\documentclass[preprint]{JHEP3}
\usepackage{epsfig}

\def\mh{m_h^{}}
\def\gev{\rm GeV}

\def\fbi{\rm fb^{-1}}
\def\tautau{\tau^+\tau^-}

\def\lsim{\mathrel{\raise.3ex\hbox{$<$\kern-.75em\lower1ex\hbox{$\sim$}}}}
\def\gsim{\mathrel{\raise.3ex\hbox{$>$\kern-.75em\lower1ex\hbox{$\sim$}}}}

\newcommand{\etmiss}{\not\hskip-5truedd E_{T}}
\title {$gg \to h \to \tautau$ at the Upgraded Fermilab Tevatron}
\author{Alexander Belyaev$^{1,2}$, Tao Han$^3$ and Rogerio Rosenfeld$^4$ 
\\
$^1$ {\it  Department of Physics, Florida State University, FL , USA}\\
$^2$ {\it Skobeltsyn Institute for Nuclear Physics,
Moscow State University, 
\\ 119 899, Moscow, Russian Federation}\\
$^3$ {\it Department of Physics, University of Wisconsin,
Madison, WI 53706, USA}\\
$^4$ {\it Instituto de F\'\i sica Te\' orica, Universidade 
Estadual Paulista,\\
Rua Pamplona 145, 01405-900 - S\~ao Paulo, S.P., Brazil}
}

\abstract{ 
We study the neutral Higgs boson production via the gluon
fusion process with the $\tautau$ final state
at the upgraded Fermilab Tevatron, including a complete simulation 
of signal channels and leading background processes.
For the SM Higgs boson, this $h\to \tautau$ channel
may provide important addition for the
Higgs boson discovery in the mass range $120-140$ GeV.
In minimal supersymmetric models, natural enhancement for
the signal rate over the SM expectation makes the 
$h,H,A\to \tautau$ signal observable for large $\tan\beta$
and low $M_A$, which may lead to full coverage for SUSY Higgs 
parameters at the Tevatron with a moderate integrated luminosity.
}


\begin{document}  

It has been extensively discussed to what extent the Higgs
bosons can be discovered at the upgraded Tevatron \cite{Run2}. 
The leading contribution to the Higgs boson production at 
hadron colliders comes from gluon fusion via heavy quark
loops, with a typical cross section of one pb at the 
Tevatron energies. However, a light Higgs boson 
mainly decays into $b \bar{b}$ and the huge QCD background  
precludes any hope for finding the Higgs boson in this channel. 
That is the reason why the most favorable
process for finding the Higgs boson at the Tevatron up to 130 GeV 
is the associated production ${W,Z} + h$ \cite{wh}, 
with a cross section of the order of $0.5$ pb.   
On the other hand, for a SM Higgs boson of mass 140 GeV 
$<m_h<190$ GeV, it may be possible to observe the mode
$h\to WW^*$ \cite{ww} if higher luminosity of order 
$10-30\ \fbi$ becomes available.

In this paper we study the feasibility of utilizing
the gluon fusion process but with the $\tautau$ final state
\vskip -0.2cm
\begin{equation}
p\bar p \to gg\to h\to \tau^+ \tau^-.
\label{htautau}
\end{equation}
There are several motivations to study this channel. First,
it is very difficult at the Tevatron to find a SM
Higgs boson signal in the mass range $130-140$ GeV at
the interplay between $b\bar b$ and $WW^*$ final states. 
It would be desirable to find other potentially useful
channels \cite{dave} and the $\tautau$ mode is 
a natural candidate to consider
since it has a branching ratio of about 
$10\% $ in the
mass range of interest. Second, this mass range of the
current interest is the probable region 
for $m_h$ in Supersymmetric models (MSSM) 
and one may expect some possible 
enhancement for process Eq.~(\ref{htautau}) due
to sparticle loops \cite{squark-loop,abd,h2t-mc}.
Thirdly, it is especially important to determine the relative coupling 
strength of $h\tautau$ and $hb\bar b$ since many new physics
scenarios predict different relations of these Yukawa
couplings \cite{taumodel}.
Finally, since $\tau$ leptons are a prominent signal 
for various models of new physics, 
the final state containing $\tau^\pm$'s has been
considered in many recent studies \cite{taus}, and
their experimental identification has become better 
understood \cite{taujets}. 
We find that the channel of Eq.~(\ref{htautau})
may provide important addition for the SM
Higgs boson discovery in the mass range $120-140$ GeV.
Especially in the MSSM, significant improvement 
at low $M_A$ and high $\tan\beta$ may lead to full 
coverage for SUSY Higgs parameters at the Tevatron.

In order to perform a complete signal and background simulation, 
we use the event generator PYTHIA v6.134 \cite{PYTHIA}
for both signal and backgrounds. The effects of initial and final 
state radiation (IFSR) and
hadronization have been taken into account. 
Final state decays of the polarized $\tau$ leptons have been 
treated properly by making use of the package
TAUOLA v2.6 \cite{tauola}.
CTEQ4M parton distributions \cite{cteq} have been used. 
The QCD scale $Q$ for both the factorization and renormalization
is set to the average of transverse momenta 
of the outgoing particles at the parton level.
Because of the limited statistics, we
consider all possible $\tau$-decay modes, hadronic and leptonic. 
Detector parameters and energy resolution were chosen the same 
as in simulation for the Run II Workshop~\cite{Run2}.
In particular, the jet energies are smeared according to
a Gaussian spread 
$${\Delta E\over E}={0.8\over {\sqrt{E}\ \gev}}.$$

Due to the missing energy carried away by the two neutrinos 
$\nu_\tau \bar\nu_{\tau}$ in the final state, we can only effectively 
reconstruct the Higgs boson mass from the decay
products of $\tautau$ if the tau pairs are not back--to--back
in the transverse plane \cite{ellisetal}. 
For this reason we need to consider the Higgs boson production 
with a finite transverse momentum ($p_T^h$). 
We adopt the process $gg(q)\to hg(q)$+initial-final 
state radiation (IFSR) for the signal simulation
as implemented in \cite{PYTHIA}.
The cross section of this process is of the order of 0.3 pb 
for $m_h$= 120 GeV with the cut $p_T^h >20\ \gev$.
With the branching fraction for $h \to \tautau$ of 
roughly $10\% $, one obtains cross sections for $p_T^j>20$ GeV
at $\sqrt s=2$ TeV 
\begin{equation} 
\sigma(p \bar{p} \to h j\to \tautau j) = 44,\ 28,\ 15\ {\rm fb \ for}\ 
m_h = 120,\ 130,\ 140\ {\gev}, 
\end{equation}
respectively.

The leading irreducible background to the signal is
$
p\bar{p}\to Zj\to \tautau j.
$
The other reducible but huge background comes from
QCD jets that can fake a $\tau$ final state,
$
p\bar{p}\to jjj\to \tautau j
$.
Using the acceptance cuts on $p_T^j$, pseudo-rapidity and
separation of the final state jets 
\begin{equation}
p_T^j > 20\ \gev,\quad |\eta_j|<2,\quad \Delta R_{jj}>0.5,
\label{accept}
\end{equation}
the cross sections for the backgrounds are
\begin{eqnarray}
  \sigma(p\bar{p}\to Zj \to \tautau j) &=&  7\times 10^4 {\rm fb},\\ 
  \sigma(p\bar{p}\to jjj) &=& 2.5\times 10^8 {\rm fb}.
\end{eqnarray}
The overwhelming background is formidable and a more efficient set 
of kinematical cuts is necessary in order to have a chance of 
extracting the signal. 

Although less important, there are other backgrounds 
that we should comment on. 
First, the final state $W^\pm jj$ with $W^\pm \to e/\mu/\tau$
and $j\to \tau$ can constitute a background to the signal.
The production rate with the jet cuts of Eq.~(\ref{accept})
is $\sigma(W^\pm jj)=300$ pb. With $W$ leptonic decay and
a jet faking a $\tau$, this background rate becomes about 200 fb.
The next potentially sizeable background is from $W^+W^-j$
production. This background has a rate $2.0$ pb. 
With $W$ leptonic decay it becomes  
$\sigma(W^+W^-j\to \ell^+\ell^- j)=220$ fb ($\ell=e,\mu,\tau$).
These background rates of $Wjj$ and $WWj$ are still somewhat
larger than the signal rate. However, since those backgrounds 
are continuous ones, they will not be important after the 
judicial cuts and especially after the Higgs mass reconstruction,
as we will discuss next.

To unambiguously identify the Higgs boson signal, one must 
reconstruct the mass peak $M_{\tau\tau}$ at $m_h$. 
This is also the most efficient way to discriminate against the
backgrounds. Due to the fact that the $\tautau$ from Higgs decay are 
ultra-relativistic, the jet (or lepton) and the neutrino(s)
from the $\tau$ decay are essentially collinear along 
${\vec p}_{\tau}$ \cite{ellisetal}. 
We can thus solve for the two neutrino momenta
as long as the $\tautau$ are not collinear.
Alternatively, one could consider
to make use of the cluster transverse mass variable
$M^T_{\tau \tau}$ \cite{tm}. This transverse mass
variable should reach maximum near $m_h$, but has a broad
tail below $m_h$.

Depending on the $\tau^\pm$ decay modes, events with $\tautau j$ 
signature lead to final states such as:
the pencil-like two $\tau$-jets +$j$; 
or one pencil-like $\tau$-jet+ lepton $+j$; 
or two leptons + $j$. All decay channels 
have at least two missing neutrinos 
$\nu_{\tau}\bar\nu_{\tau}$, and each charged lepton 
$\ell$ will be accompanied by another neutrino $\nu_\ell$. 
Among those channels, the leptonic channels lead to a better
energy-momentum determination but have a smaller branching
fraction. On the other hand, the hadronic channels have a
higher rate but a poor energy resolution for mass reconstruction.

\FIGURE[tb]{\centerline{
            \hspace*{-0.7cm}\epsfxsize=0.55\hsize \epsffile{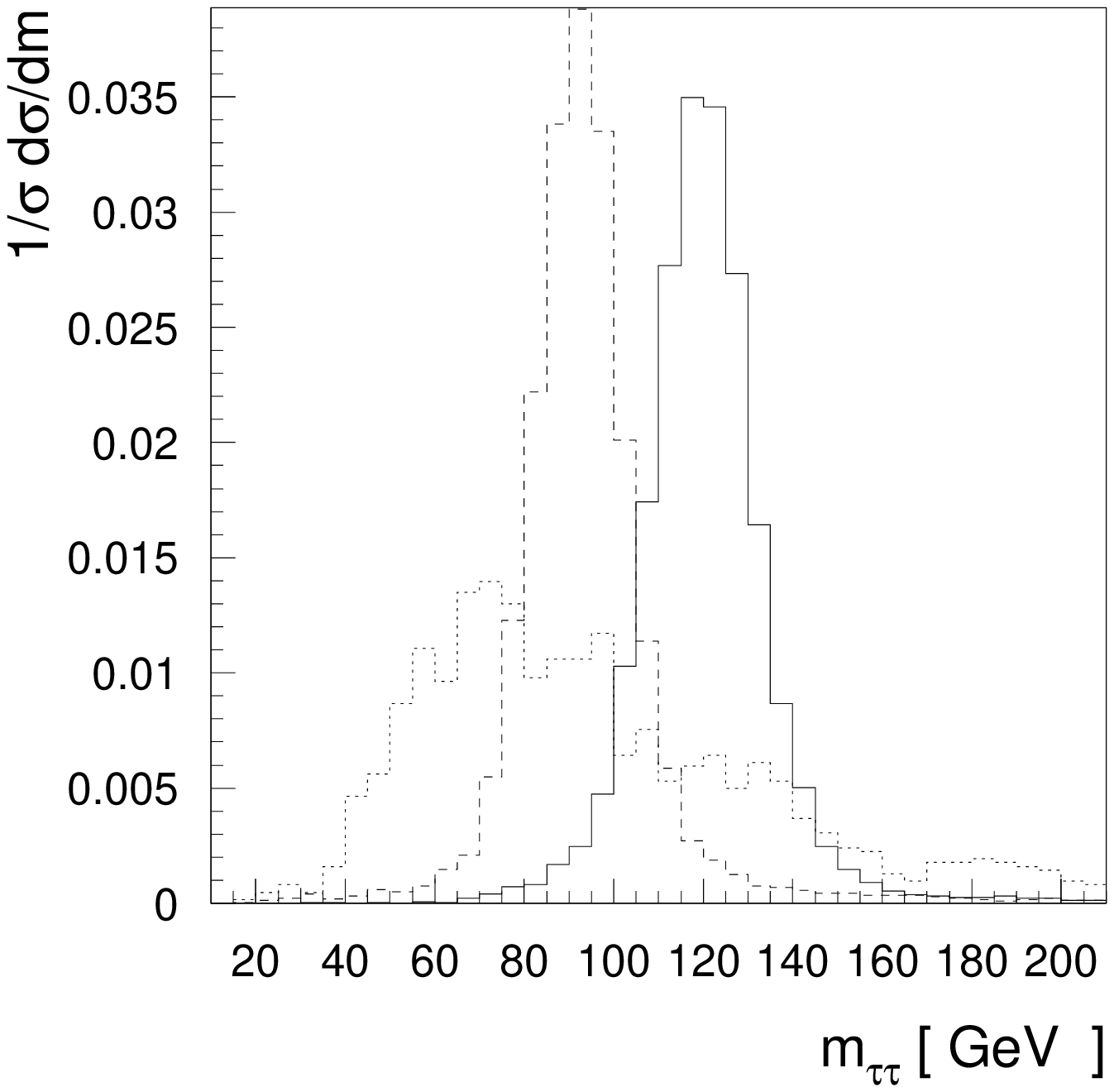}
	    \hspace*{-0.7cm}\epsfxsize=0.55\hsize \epsffile{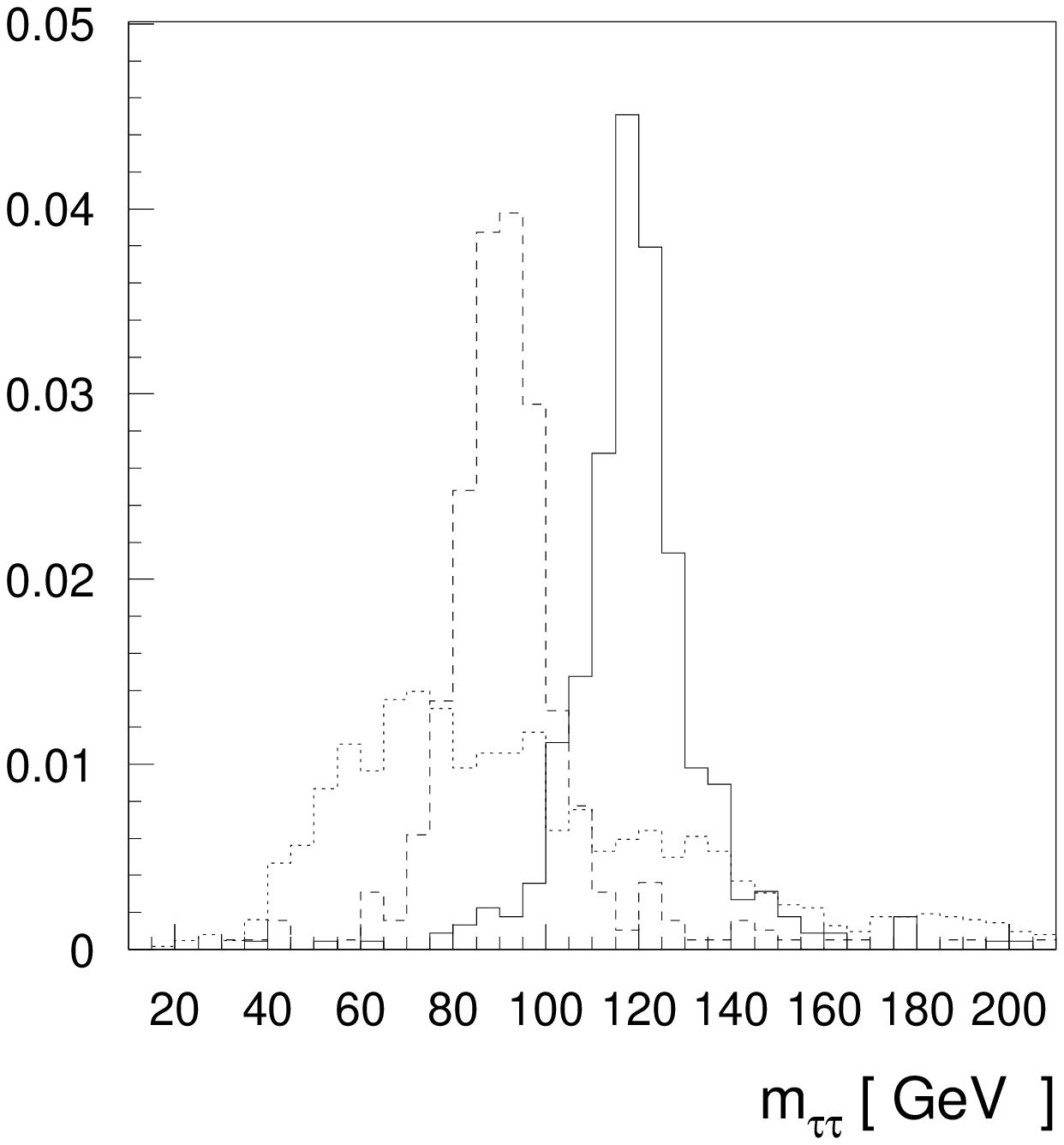}}
	    \vspace*{-0.7cm}
\caption{\label{fig:mtau} Normalized reconstructed mass $M_{\tau \tau}$ 
distributions for $\tautau$ decaying (a) hadronically 
and (b) leptonically.
The solid lines are for $m_h=120$ GeV, 
the dashed lines for the $Zj$ background and the 
dotted lines for $jjj$ background. }
}

In Fig.~\ref{fig:mtau} we show the normalized reconstructed 
mass distributions of $M_{\tau \tau}$
for the signal and backgrounds for $m_h = 120$ GeV. 
Figures \ref{fig:mtau}(a) is for both $\tautau$ decaying 
hadronically and Fig.~\ref{fig:mtau}(b)
for $\tautau$ decaying leptonically. Although the $jjj$
background rate is a lot higher than that of $Zj$ to begin
with, the reconstructed mass spectra are sufficiently
different from the peak structure of the signal. This
feature is also true for other faked backgrounds $Wjj$ and $WWj$.
In contrast, the $Zj$ background naturally presents a peak at 
the $Z$ mass and leads to a long tail after $M_Z$. This 
constitutes the major irreducible background as we will
see later. This becomes the limiting factor for us to explore
a Higgs boson below and near 110 GeV.
As anticipated, the leptonic channels have
sharper signal mass peaks. 
Taking into account the kinematical features of the 
signal and backgrounds discussed above, we devise 
the following set of kinematical cuts. We first require 
the $\tau$ identification for the hadronic ($\tau_j$) or leptonic
($\tau_\ell$) decay, Cut~I:
\begin{eqnarray}
\tau_j:\ &&\bullet\  p^j_T > 15\ \gev,|\eta^j|<2 \nonumber\\
&&\bullet\ \mbox{one or three tracks in $10^\circ$ cone} 
\ \      \mbox{with no additional tracks 
in $30^\circ$ cone}\nonumber\\
&&\bullet\ \mbox{the invariant mass of tracks is less than 2 \gev} \nonumber\\
\tau_\ell:\ &&\bullet\ p^\ell_T > 10\ \gev \ |\eta^\ell| < 2; \nonumber\\
&&\bullet\ \mbox{no additional tracks in $30^\circ$ cone}
\ \ {\rm for\ jjj\ mode:}\ \etmiss > 20 \ {\gev}, 
\nonumber
\end{eqnarray}
where the missing transverse energy $\etmiss$ is defined by the
imbalance of the observed particles, and it is also smeared according
to Gaussian distribution with standard deviation $0.5 \sqrt \etmiss$. 
The $\etmiss$  cut is to help triggering the jjj events.
For the $jj$ channel we require $2\tau_j$ with no 
isolated electrons or muons (defined by $\tau_\ell$),
for the $j\ell$ channel we require only one $\tau_j$  
and one isolated lepton, and for the $\ell\ell$ channel 
we require no  $\tau_j$ in addition to the
two isolated leptons.

In Cut~I, the cut on the jet invariant mass is essential 
for reducing the huge QCD background in which a QCD parton
jet fakes a $\tau$. The efficiency of $\tau$ ID 
from the $Z$ decay is in agreement with more realistic 
studies \cite{taujets}. 
The efficiency of more energetic $\tau$'s from Higgs decay has not been 
previously studied for the Tevatron with the full detector simulation.
We find that the efficiency of $\tau$ ID is quite $p_T$ and process-dependent, 
typically being  $60-70\%$ for $\tau_j$ and  $50-60\%$  for $\tau_l$
for $\mh=120-140$ GeV.
After the $\tau$-pair identification, we then impose the next level 
of refined kinematical cuts, Cut II:
\begin{eqnarray}
&&\bullet\  p^{\tau_j}_T > m_h/6,\qquad  p^{\tau_e}_T > m_h/9,\nonumber \\
&&\bullet\  \mbox{additional jet (not $\tau_j$) with   
$p^{j}_T > m_h/6$} \ \ \mbox{and  $|\eta^j|<4$.}
\nonumber
\end{eqnarray}

Our final set of cuts defines the optimal windows in $M_{\tau \tau}$
and $M^T_{\tau \tau}$ that will be used to confirm the existence
of a signal as well as to estimate the significance
for the signal observation, Cut~III(A):
\begin{eqnarray}
&& m_h - 0.5 \Delta M  < M_{\tau \tau} < m_h+3 \Delta M, 
\nonumber\\
&& m_h -  4 \Delta M < M^T_{\tau \tau}  <  m_h+3 \Delta M,\nonumber
\end{eqnarray}
for pure hadronic decay of both $\tau$'s and Cut~III(B):
\begin{eqnarray}
&& m_h -       \Delta M  < M_{\tau \tau}  <  m_h+3 \Delta M,\nonumber \\
&& m_h -  4.5 \Delta M < M^T_{\tau \tau}  <  m_h+3 \Delta M,\nonumber
\end{eqnarray}
for others channels of  $\tau$ decay,
where $\Delta M = \sqrt{m_h/\mbox{GeV}}$ GeV.

In the case of the QCD faked background to $\tau$,
we do not directly go through the procedure outlined in
Cut~I as for the signal. Instead, we use the probability 
for a jet to be misidentified as a $\tau_j$ to be 
 $0.5\%$ \cite{taus}, and $\tau_l$ to be
$0.01 \%$. We then apply Cuts II and III. 
Since the probability for QCD background 
to fake two leptons is very small, the
background to the di-lepton channels 
is negligible in comparison with the signal,
especially with the irreducible $Zj$ background.

With the substantial efforts discussed above, 
we have effectively suppressed the backgrounds with respect to the signal.
The signal efficiencies are at a percentage level.
Table \ref{events} shows the final number of events for
all signal channels and their corresponding backgrounds
at the Tevatron with an integrated luminosity 10 fb$^{-1}$.
Also shown in Table \ref{events} are the signal-to-background 
ratios (S/B).
Because of the overwhelming QCD background, the S/B ratio  is the highest for 
$\ell\ell$ channel of $\tau$ decay   ($\sim 6-8\%$), 
intermediate for     $j\ell$ channel ($\sim 4-5\%$),
and the lowest  for $jj$ channel     ($\sim 3\%$).
With respect to $S/\sqrt{B}$ and $S/{B}$ ratios,  
the $j\ell$ channel is probably the best one for signal identification.
Nevertheless, the rather small $S/B$ ratios
render the signal observation systematically challenging.

\TABLE[h]{
\begin{tabular}{| r |c c c c| c c c c|c c c c|}   
\multicolumn{1}{|c|}{$m_h$}&
\multicolumn{4}{c|}{$120$~GeV} &
\multicolumn{4}{c|}{$130$~GeV} &
\multicolumn{4}{c|}{$140$~GeV} \\
\hline
channels  &
 $hj$ & $Zj$ &  $jjj$ & S/B(\%) &
 $hj$ & $Zj$ &  $jjj$ & S/B(\%) &
 $hj$ & $Zj$ &  $jjj$ &  S/B(\%) \\ 
$jj    $&32  &713& 559  & 2.5 &20   &    281 &    346 & 3.2 &   10   &    164  &     195 &  2.7\\
$j\mu  $&18  &430&  13  & 4.1 &10   &    137 &     8.5& 6.9 &   5.3  &    67   &     5.3 &  7.3\\
$je    $&17  &338&  13  & 4.8 &10   &    159 &     8.5& 6.0 &    5.0 &    52   &     5.3 &  8.7\\
$\mu\mu$& 1.4& 18&  0.26& 7.7 & 0.85&     10 &    0.17& 8.4 &   0.39 &	  6.0  &     0.11&  4.9\\
$ee    $& 1.2& 18&  0.26& 6.5 & 0.62&     10 &    0.17& 6.1 &   0.31 &	  6.0  &    0.11 &  5.1\\
$\mu e $& 2.5& 40&  0.26& 6.2 & 1.5 &     24 &    0.17& 5.1 &   0.78 &	  15   &    0.11 &  5.2
\end{tabular}
\caption{\label{events}Final number of events of signal for all 
channels from $h\to \tautau$ for representative Higgs boson masses,
corresponding backgrounds at $\sqrt s=2$ TeV per 10 fb$^{-1}$
integrated luminosity and S/B ratio.}
}

Our main results, the luminosities required 
for a $95\%$ CL exclusion or a $5\sigma$ discovery 
of the SM Higgs boson via $h\to \tautau$ at $\sqrt s=2$ 
TeV for $m_h = 120,\ 130$ and $140$ GeV,
are summarized in Table \ref{tabsinal}. 
A $95\%$ CL exclusion limit 
for the SM Higgs boson in the mass range 
$120-140$ GeV via this single channel $h\to \tautau$ 
would require a total integrated luminosity
of $14-32$ fb$^{-1}$. To gain an idea on the signal observation
in theories beyond the SM, 
also given in Table \ref{tabsinal} are the necessary enhancement 
factors $\kappa$ over the SM rate ($\sigma_{new}=\kappa \sigma_{SM}$) 
for reaching $95\%$
and 5$\sigma$ signal with 2 and 15 fb$^{-1}$ luminosity.

\TABLE[h]{
\begin{tabular}{| r |c c c | c c c|c c c|}   
\multicolumn{1}{|c|}{$m_h$}&
\multicolumn{3}{c|}{$120$~GeV}&
\multicolumn{3}{c|}{$130$~GeV}&
\multicolumn{3}{c|}{$140$~GeV}\\ \hline
 $95\%$ CL exclusion  $L$(fb$^{-1}$) & 
 \multicolumn{3}{c|}{  14 } &  
 \multicolumn{3}{c|}{  18 } &  
 \multicolumn{3}{c|}{  32 } \\ \hline
 $3\sigma$ discovery $L$(fb$^{-1}$) & 
 \multicolumn{3}{c|}{33 } &  
 \multicolumn{3}{c|}{42 } &  
 \multicolumn{3}{c|}{77 } \\ \hline 
 $5\sigma$ discovery $L$(fb$^{-1}$) & 
 \multicolumn{3}{c|}{ 93 } &  
 \multicolumn{3}{c|}{120 } &  
 \multicolumn{3}{c|}{210 } \\ \hline  \hline
 $\kappa$ for 95\% CL (2 fb$^{-1}$)  &
 \multicolumn{3}{c|}{  2.7} &  
 \multicolumn{3}{c|}{  3.0} &  
 \multicolumn{3}{c|}{  4.0} \\
 $\kappa$ for 95\% CL (15 fb$^{-1}$)  &
 \multicolumn{3}{c|}{  0.97} &  
 \multicolumn{3}{c|}{  1.1} &  
 \multicolumn{3}{c|}{  1.5} \\
 $\kappa$ for 3$\sigma$ (2 fb$^{-1}$)  &
 \multicolumn{3}{c|}{  4.1 } &  
 \multicolumn{3}{c|}{  4.6 } &  
 \multicolumn{3}{c|}{  6.2 } \\
 $\kappa$ for 3$\sigma$ (15 fb$^{-1}$)  &
 \multicolumn{3}{c|}{  1.5 } &  
 \multicolumn{3}{c|}{  1.7 } &  
 \multicolumn{3}{c|}{  2.3 } \\
 $\kappa$ for 5$\sigma$ (2 fb$^{-1}$)  &
 \multicolumn{3}{c|}{  6.8 } &  
 \multicolumn{3}{c|}{  7.7 } &  
 \multicolumn{3}{c|}{  10 } \\
 $\kappa$ for 5$\sigma$ (15 fb$^{-1}$)  &
 \multicolumn{3}{c|}{  2.5 } &  
 \multicolumn{3}{c|}{  2.8 } &  
 \multicolumn{3}{c|}{  3.8 } \\
\end{tabular}
\caption{\label{tabsinal} Integrated luminosities needed
to reach a $95\%$ CL exclusion, 3 and $5\sigma$ discovery
for a SM Higgs boson at the Tevatron, 
and the enhancement factor $\kappa$ (at 2 and 15 fb$^{-1}$)
needed to reach a $95\%$ CL exclusion and $5\sigma$ 
discovery.  }
}

The $\tautau$ channel
can provide new addition in combination with the other
promising channels such as $Wh, Zh$ and $h\to WW$ \cite{Run2,wh,ww} 
to improve the overall observability of the SM Higgs boson.
To illustrate the potential improvement, 
we estimate the total integrated luminosities needed for the
$95\%, 3\sigma,5\sigma$ effects for $m_h=120-140$ GeV by combining 
the Run II report \cite{Run2} and our $h\to \tautau$ results, 
according to a relation 
$$L^{-1}=L_1^{-1}+L_2^{-1},$$
as shown in Table \ref{lum}.
We have followed the convention from the Run II report 
that the numbers in Table \ref{lum} correspond to the
delivered machine luminosity; while the significance values have 
been evaluated by combining CDF and D0 (doubling the delivered
luminosity).
Especially for the most difficult region $m_h\sim 140$ GeV, 
the luminosity needed may be reduced by about 40\%
with the addition of the $h\to \tautau$ channel.

\TABLE[thb]{
\begin{tabular}{| r |c c c | c c c|c c c|}   
\multicolumn{1}{|c|}{$m_h$ (GeV)}&
\multicolumn{3}{c|}{$120$}&
\multicolumn{3}{c|}{$130$}&
\multicolumn{3}{c|}{$140$}\\ \hline
 $L$\ at\ $95\% \ (\fbi)$ & 
 \multicolumn{3}{c|}{ 1.8 (2.5) } &  
 \multicolumn{3}{c|}{ 3.2 (5) } &  
 \multicolumn{3}{c|}{ 6.6 (11) } \\ \hline
 $L$\ at\ $3\sigma\ (\fbi)$ & 
 \multicolumn{3}{c|}{ 4.4 (6) } &  
 \multicolumn{3}{c|}{ 11 (21) } &  
 \multicolumn{3}{c|}{ 15 (24) }\\  \hline
 $L$\ at\ $5\sigma\ (\fbi)$  &
 \multicolumn{3}{c|}{ 13 (18) } &  
 \multicolumn{3}{c|}{ 20 (31) } &  
 \multicolumn{3}{c|}{ 42 (70) } \\
\end{tabular}
\caption{\label{lum} Integrated luminosities needed 
(for one of the two experiments)
to reach a $95\%$ CL exclusion, 3 and $5\sigma$ discovery
for a SM Higgs boson by combining the current 
{\protect $\tautau$} results and the Run II report {\protect \cite{Run2}.}
Numbers in parentheses are from {\protect \cite{Run2}}. 
}}
\FIGURE[tb]{
\centerline
{\hspace*{0.5cm}\epsfxsize=0.62\hsize\epsffile{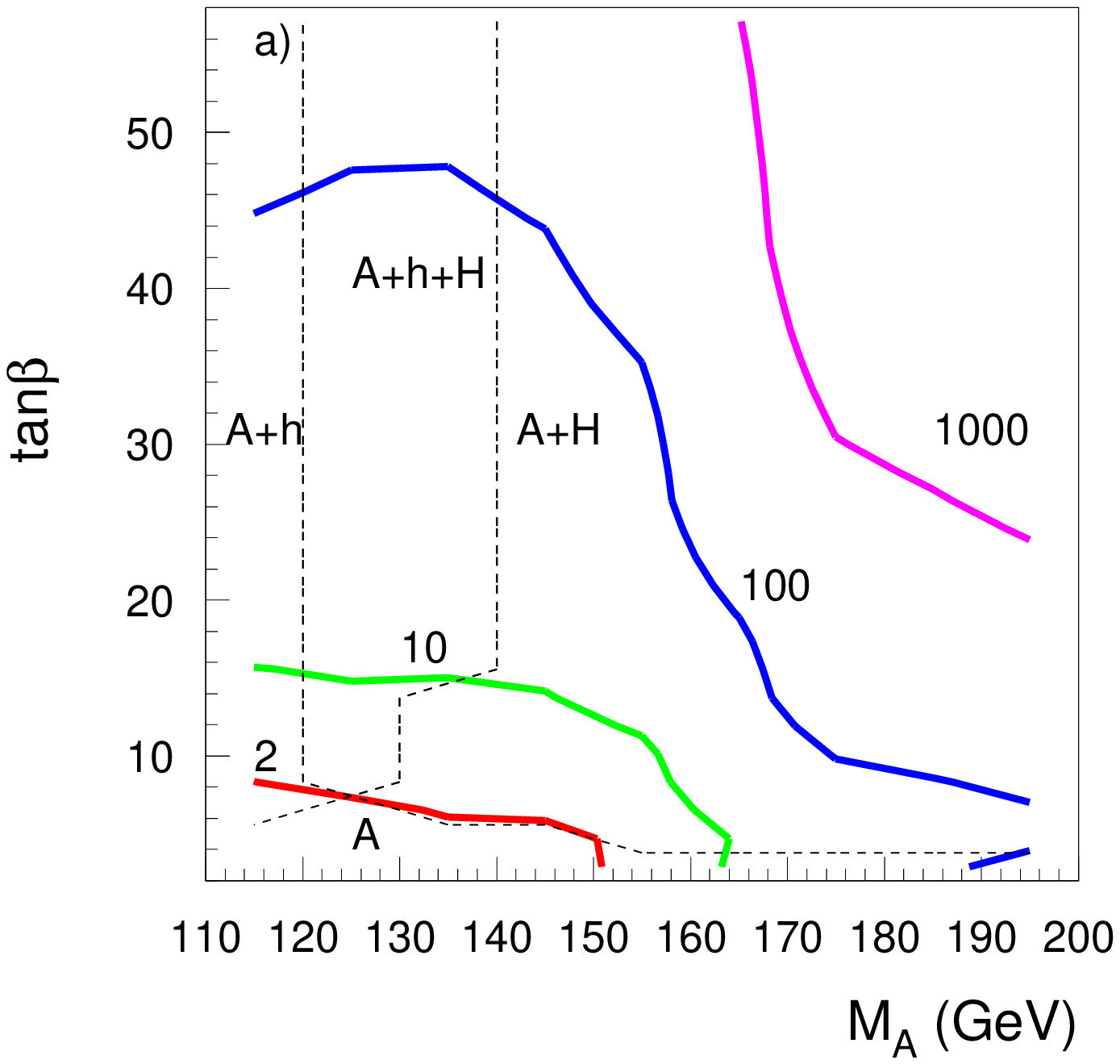}
 \hspace*{-0.5cm}
 \epsfxsize=0.66\hsize\epsffile{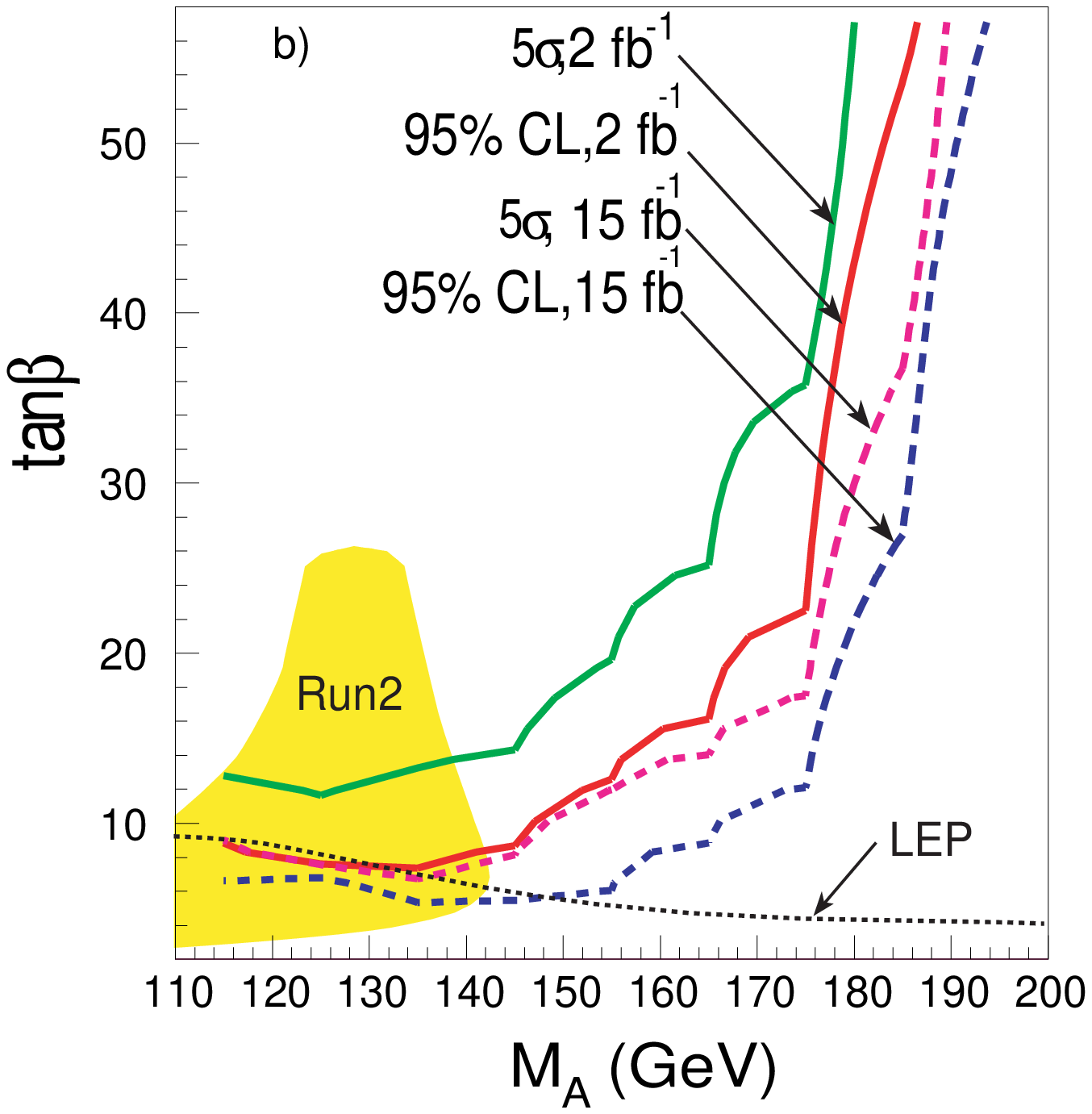}}
\caption{
\label{factor} (a) Contours (above the curves) in 
($M_A, \tan\beta$) plane for the enhancement factor 
above the SM Higgs boson signal rate as labeled. 
The contributing
Higgs states are also indicated by $A,h,H$ separated by
the dashed lines;
(b) contours (above the curves) for $95\%$ CL exclusion 
and 5$\sigma$ discovery for 
$p\bar{p}\to (A,h)+j\to \tautau j$ process within MSSM 
at the Tevatron with 2 and 15 $\fbi$.}
}

In the minimal supersymmetric models, there are two more
neutral Higgs bosons $H,A$ to contribute to the $\tautau$
mode. In addition, there are new squark-loop contributions 
to Higgs boson production through gluon fusion.
In our analyses, we use HIGLU program \cite{higlu} 
for the calculation of the cross section of 
Higgs production process in the MSSM.
For simplicity, we fix the squark masses to 1 TeV. We find
that left-right squark mixing is not a crucial factor in
our analysis. 
This simplification allows us to illustrate the potential of
the Tevatron search by the Higgs sector
parameters $M_A-\tan\beta$ in the conventional way 
in a two dimensional plane. In Fig.~\ref{factor}(a)  
we plot contours of the enhancement factor over the
SM expectation which is defined to be the rate
at $m_h^{SM}=M_A$. The large enhancement above the SM mainly 
comes from the new channels from $H,A$.
The contributing Higgs states are also indicated by
the dashed lines in Fig.~\ref{factor}(a) with the explicit
labels  $A,h,H$.
Also, large $\tan\beta$
will enhance the effects from bottom/sbottom loops.
On the other hand, the decay branching fractions
to $\tau^+\tau^-$ may be enhanced by about $30\%$ at most for
various choices of the SUSY parameters. For $M_A\sim 150-180$ GeV,
$H,A\to \tautau$ is still observable; while the SM mode dies
away due to the opening of the dominant $WW,ZZ$ channels.
When combing the cross sections 
from channels of $A$ and $h$ production,
we have used Gaussian combination criteria.
In Fig.~\ref{factor}(b) 
we show the $95\%$ CL exclusion and 5$\sigma$ discovery
contours in MSSM for the $\tautau$ mode at the Tevatron.
The direct searches at LEP2 \cite{lep2} have put a
{\it lower} bound on the parameters
as indicated by the dotted curve. 
The shaded region is the uncovered $5\sigma$ hole 
at Run II with 15 fb$^{-1}$ \cite{Run2,h2t-mc}
without considering the $\tautau$ mode. We see that the addition
of the $\tautau$ mode could reach 5$\sigma$ (2$\sigma$) 
full coverage for SUSY Higgs parameters
with 15 fb$^{-1}$ (2 fb$^{-1}$).
As a final remark, we note that in some region of the parameter 
space $hb\bar{b}$ Yukawa coupling is accidentally suppressed due
to radiative effects. This could lead to a substantial 
increase of the $h\to \tau\bar{\tau}$
branching fraction \cite{taumodel}. 
However, we found that this gain in branching fraction is mostly
balanced by the reduction in the Higgs production
via the $hb\bar{b}$ coupling.

In summary, we have studied the potential of the upgraded
Tevatron for searches of the neutral Higgs boson 
in the SM and in the MSSM via the $\tau\tau+j$ channel. 
We found that with this single channel alone,
the total integrated luminosity needed 
for a $95\%$ CL exclusion is 
$14-32$ fb$^{-1}$ for $m_h=120-140$ GeV. 
The addition of the $h\to \tautau$ channel could 
improve the SM Higgs observation, 
as presented in Table \ref{lum}. For the most difficult region
$m_h\sim 140$ GeV, the luminosity needed for a 5$\sigma$
signal could be reduced by about $40\%$.
However, we must note that due to the rather small 
signal-to-background ratios, the search for $\tautau$ channels is
systematically challenging in the Standard Model case.

For the case of MSSM, the signal cross sections
can be enhanced by a significant factor due to the addition
of $H,A$, and due to high $\tan\beta$ or squark-loop contributions.
We found 5$\sigma$ (2$\sigma$) full coverage for the 
SUSY Higgs parameters with 15 fb$^{-1}$ (2 fb$^{-1}$)
by including the $\tautau$ channel.
Our analyses is applicable for other generic neutral scalar
or pseudo-scalar via gluon fusion production
and with substantial branching fraction to $\tautau$.
More details of our analyses will be presented in an extended 
version of this work \cite{us}.

\vskip 0.1cm
{\it Acknowledgments:}\ 
We thank M.~Carena, J.~Conway, Y.~Gershtein and S.~Mrenna 
for discussions, C.~Wagner and D.~Morrissey for pointing out
an inconsistency between Tables II and III in the earlier version.
This work was supported in part by Conselho Nacional de 
Desenvolvimento Cient\'{\i}fico e Tecnol\'ogico (CNPq), by
Funda\c{c}\~ao de Amparo \`a Pesquisa do Estado de S\~ao Paulo 
(FAPESP), by Programa de Apoio a N\'ucleos de Excel\^encia (PRONEX),
and in part by  US DOE grants  DE-FG02-95ER40896 and DE-FG02-97ER41022 
and in part by the 
Wisconsin Alumni Research Foundation.

\end{document}